\documentclass[reprint,superscriptaddress,showpacs,amsmath,amssymb,floatfix,aps,pdftex]{revtex4-1}

\usepackage[hiresbb]{graphicx}
\usepackage{dcolumn}
\usepackage{bm}

\usepackage{color}

\begin{document}
\title{
Continuum Excitation and Pseudospin Wave in Quantum Spin-Liquid and Quadrupole Ordered States of Tb$_{2+x}$Ti$_{2-x}$O$_{7+y}$
}

\author{Hiroaki Kadowaki}
\affiliation{Department of Physics, Tokyo Metropolitan University, Hachioji-shi, Tokyo 192-0397, Japan}

\author{Mika Wakita}
\affiliation{Department of Physics, Tokyo Metropolitan University, Hachioji-shi, Tokyo 192-0397, Japan}

\author{Bj\"{o}rn F{\aa}k}
\affiliation{Institute Laue Langevin, BP156, F-38042 Grenoble, France}

\author{Jacques Ollivier}
\affiliation{Institute Laue Langevin, BP156, F-38042 Grenoble, France}

\author{Seiko Ohira-Kawamura}
\affiliation{Neutron Science Section, MLF, J-PARC Center, Shirakata, Tokai, Ibaraki 319-1195, Japan}

\author{Kenji Nakajima}
\affiliation{Neutron Science Section, MLF, J-PARC Center, Shirakata, Tokai, Ibaraki 319-1195, Japan}

\author{Hiroshi Takatsu}
\affiliation{Department of Energy and Hydrocarbon Chemistry, Graduate School of Engineering, Kyoto University, Kyoto 615-8510, Japan}

\author{Mototake Tamai}
\affiliation{Department of Physics, Tokyo Metropolitan University, Hachioji-shi, Tokyo 192-0397, Japan}

\date{\today}

\begin{abstract}
The ground states of the frustrated pyrochlore oxide Tb$_{2+x}$Ti$_{2-x}$O$_{7+y}$ 
have been studied by inelastic neutron scattering experiments. 
Three single-crystal samples are investigated; 
one shows no phase transition ($x=-0.007<x_{\text{c}}\sim -0.0025$), 
being a putative quantum spin-liquid (QSL), 
and the other two ($x=0.000, 0.003$) show electric quadrupole ordering (QO) 
below $T_{\text{c}} \sim 0.5$ K. 
The QSL sample shows continuum excitation spectra with an energy scale 0.1 meV 
as well as energy-resolution-limited (nominally) elastic scattering. 
As $x$ is increased, 
pseudospin wave of the QO state emerges from this continuum excitation, 
which agrees with that of powder samples 
and consequently verifies good $x$ control for the present single crystal samples.
\end{abstract}

\pacs{75.10.Kt, 75.40.Gb, 75.70.Tj, 78.70.Nx}
\maketitle

\section{Introduction}
Geometrically frustrated magnets have been actively studied 
in recent years \cite{Lacroix11}. 
These include classical and quantum spin systems on 
two-dimensional triangle \cite{Wannier50,Mekata1977} and kagom\'{e} \cite{Shyozi51,Shores2005} lattices, 
and three-dimensional pyrochlore-lattice systems \cite{Gardner10}. 
For classical systems, 
prototypes of which are the trianglar-lattice antiferromagnet \cite{Wannier50} 
and the spin ice \cite{Bramwell01,Castelnovo08,Kadowaki09}, 
many investigations have been performed for several decades 
using a number of theoretical and experimental techniques \cite{Lacroix11}. 
Possibilities of quantum spin liquid (QSL) states in frustrated magnets, 
which date back to the theoretical proposal of the RVB state \cite{Anderson73}, 
are recently under hot debate. 
Highly-entangled many-body wave functions without magnetic long-range order (LRO), 
anticipated in QSL states, 
provide theoretically challenging problems \cite{Savary2017}. 
Experimentally, finding out real QSL substances, e.g., 
Refs.~\cite{Hirakawa1985,Shimizu2003,Nakatsuji2006,Li2015,Sibille2016}, 
and investigating QSL states using available techniques, e.g., 
Refs.~\cite{Han2012,Ross11,Chang12,Shen2016,Fak2017}, 
have been challenging explorations. 

A non-Kramers pyrochlore system Tb$_{2}$Ti$_{2}$O$_{7}$ (TTO) 
has attracted much attention 
since interesting reports of absence of magnetic LRO 
down to $0.1 \sim 0.4$ K \cite{Gardner99,Yasui2002,Gardner03}, 
which could be interpreted as a QSL candidate \cite{Gingras00,Kao03} 
or quantum spin ice (QSI) \cite{Molavian07,Hermele04}. 
On the other hand, a phase transition at $T_{\text{c}} \sim 0.5$ K 
detected by a specific heat peak suggesting a hidden LRO \cite{Hamaguchi04,Takatsu12}, 
seemed to contradict with the QSL interpretation. 
We resolved this contradiction by showing that ground states of TTO are highly sensitive 
to off-stoichiometry, i.e., $x$ (and/or $y$) of Tb$_{2+x}$Ti$_{2-x}$O$_{7+y}$ \cite{Taniguchi13}. 
It was shown that $x$ of powder samples is much easier to control than crystal samples, 
and that there are two ground states: 
a hidden LRO ($x > x_{\text{c}}\sim -0.0025$) state 
and a QSL ($x < x_{\text{c}}$) state \cite{Taniguchi13}. 

By carefully analyzing experimental data of TTO samples, 
we proposed an effective pseudospin-1/2 Hamiltonian 
for a typical TTO sample with the LRO groud state, 
Tb$_{2.005}$Ti$_{1.995}$O$_{7+y}$ ($T_{\text{c}} \sim 0.5$ K) 
\cite{Takatsu2016prl,Kadowaki2015,Takatsu2016JPCS,Takatsu2017}. 
This Hamiltonian consists of interaction terms among magnetic dipole moments 
and those among electric quadrupole moments \cite{Onoda10,Onoda11}. 
It naturally explains that 
the hidden LRO is an electric quadrupole order (QO) \cite{Onoda10,Onoda11}, 
and that a neighboring phase to this QO phase 
is a theoretically proposed U(1) QSL state \cite{Lee12,Hermele04}, 
possibly occurring in TTO samples without LRO \cite{Takatsu2016prl}. 

Therefore, the long-standing question of ``what is the QSL state of TTO?'' 
\cite{Gingras00,Gingras14} is now about to be 
addressed using well-controlled single-crystals \cite{Wakita2016}. 
In other words, one needs to pickup useful clues for solving this question 
from many experimental results reported to date, e.g., 
Refs.~\cite{Gardner99,Yasui2002,Gardner03,Gingras00,Hamaguchi04,Mirebeau07,Bertin12,Chapuis2010,Yaouanc2011,Lhotel2012,Guitteny2013,Bonville11,Petit12,Bonville2014,Fennell2012,Fennell14,Gaulin2011,Fritsch13,Fritsch14,Zhang14,Princep2015,Bovo2017,
Hirschberger2015,Ruminy2016,Ruminy2016L,Constable2017,Kermarrec2015,Ruminy2016S}, 
which could suffer from the off-stoichiometry and its inhomogeneity problems to a certain extent. 
Although the sample stoichiometry does not necessarily affect all experimental facts, 
one should cautiously reconsider results of these references. 
In particular, it should be noted that our evaluation of the sample stoichiometry and 
those of Refs.~\cite{Guitteny2015,Kermarrec2015,Ruminy2016S} 
likely do not coincide. 

In this work, we have prepared large single-crystal samples of Tb$_{2+x}$Ti$_{2-x}$O$_{7+y}$ 
with controlled $x$ ($y$ is determined by the oxidation condition) \cite{Wakita2016}, 
and performed inelastic neutron scattering (INS) experiments. 
The resulting INS spectra of TTO samples with and without $T_{\text{c}}$ 
show that small variation of $x$ really induces a continuous change of 
INS spectra at low temperatures, 
reflecting the QSL and QO ground states. 
This agrees with our previous results of INS on powder samples \cite{Taniguchi13,Takatsu2016prl}. 

\section{Experimental Methods}
Polycrystalline samples of Tb$_{2+x}$Ti$_{2-x}$O$_{7+y}$ were 
prepared by the standard solid-state reaction \cite{Taniguchi13}. 
The two starting materials, Tb$_{4}$O$_{7}$ and TiO$_2$, 
were heated in air at 1350~$^\circ$C for several days with 
periodic grindings to ensure a complete reaction. 
The value of $x$ was adjusted by changing the mass ratio 
of the two materials. 
Single crystal Tb$_{2+x}$Ti$_{2-x}$O$_{7+y}$ rods were 
grown by the floating zone (FZ) technique 
from feed rods of sintered powder samples with $-0.04<x<-0.002$ \cite{Wakita2016}. 
Crystal growth was carried out in an Ar gas flow atmosphere 
using a double ellipsoidal image furnace (NEC SC-N35HD).  
They were post-annealed for 2-7 days at 1000~$^{\circ}$C in air. 
The $y$ value was determined by this oxidizing condition, where 
an unmeasurably small increase of $y$ occurred. 
The off-stoichiometry parameter $x$ was evaluated 
by measuring the lattice parameter $a(T,x)$ 
using a high-resolution X-ray diffractometer (RIGAKU SmartLab) 
equipped with a Cu K$_{\alpha 1}$ monochromator \cite{Wakita2016}.
Details of this evaluation are described in the appendix. 

For INS experiments 
we chose seven 15-20 mm crystal rods with small concentration gradient 
cut from longer crystal rods with $\sim 2$ mm in diameter, 
and assembled three samples. 
One sample is $x = -0.007$ ($< x_{\text{c}}$, in the 
QSL range) and consists of three 15 mm crystal rods with 1.2 g in total weight.
In the inset photograph of Fig.~\ref{theta2theta_scan} we show 
one 15 mm crystal of this QSL sample. 
To evaluate $x$ of this crystal, 
two samples ($\sim 1$ mm) for the X-ray measurement 
were cut from two neighbors being very close to the both ends. 
The resulting lattice parameters ensure that 
$x=-0.007 \pm 0.002$. 
Another crystal sample for INS is $x = 0.000 \pm 0.002$ ($> x_{\text{c}}$, in the 
QO range) and consists of three 15 mm crystals with 1.1 g in total weight.
These QSL and QO samples are co-aligned within 1.5 degrees. 
The third crystal sample is a 20 mm rod with $x = 0.003 \pm 0.002$, which is in the QO range. 
It is 0.6 g in weight. 
\begin{figure}[htb]
\centering
\includegraphics[width=8.0cm,clip]{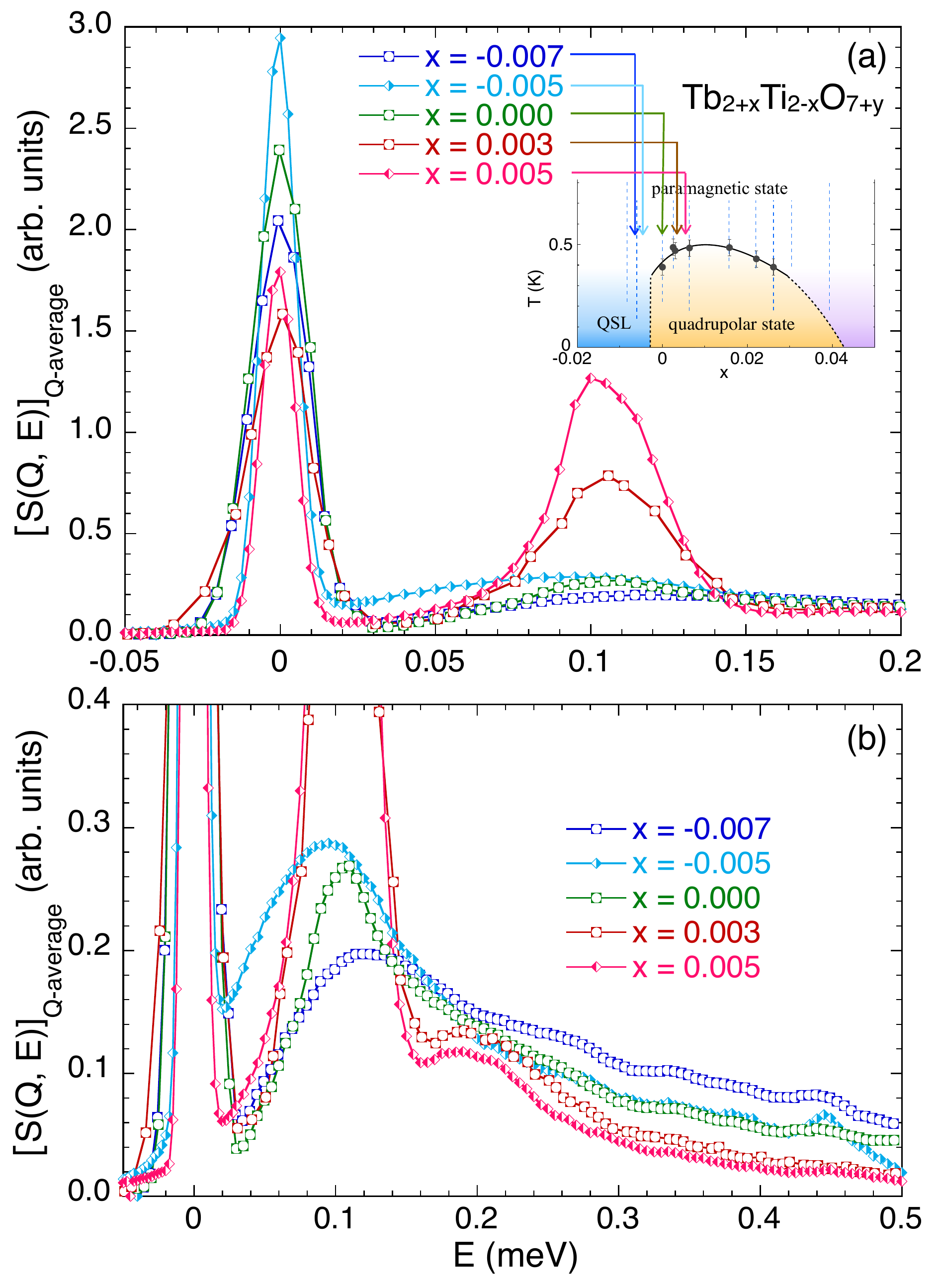}
\caption{(Color online) 
Energy spectra $S(\bm{Q},E)$ averaged in a wide $\bm{Q}$-range, 
$\bm{Q}=(h+k,h-k,l)$ with $0<h<0.9$, $0.75<l<1.75$, $-0.25<k<0.25$ (r.l.u), 
at $T=0.1$ K for $x = -0.007$, 0.000, and 0.003 crystal samples. 
Two scales are used to show (a) a low energy part and (b) a low intensity part.
These spectra are plotted with previous results of powder samples 
with $x = -0.005$ and 0.005 taken at $T=0.1$ K \cite{Taniguchi13,Takatsu2016prl}.
The energy spectra of the powder samples are averaged in 
$0.3 < |\bm{Q}| < 0.9$ {\AA}$^{-1}$. 
In the inset of (a), these samples are shown by arrows 
in the $x$-$T$ phase diagram of Ref.~\cite{Wakita2016}. 
The intensity scales of the five samples are approximately normalized 
as described in the text. 
}
\label{Escan_Qav_0p1K_PhaseDiagram}
\end{figure}
\begin{figure*}[hbt]
\centering
\includegraphics[width=17cm,clip]{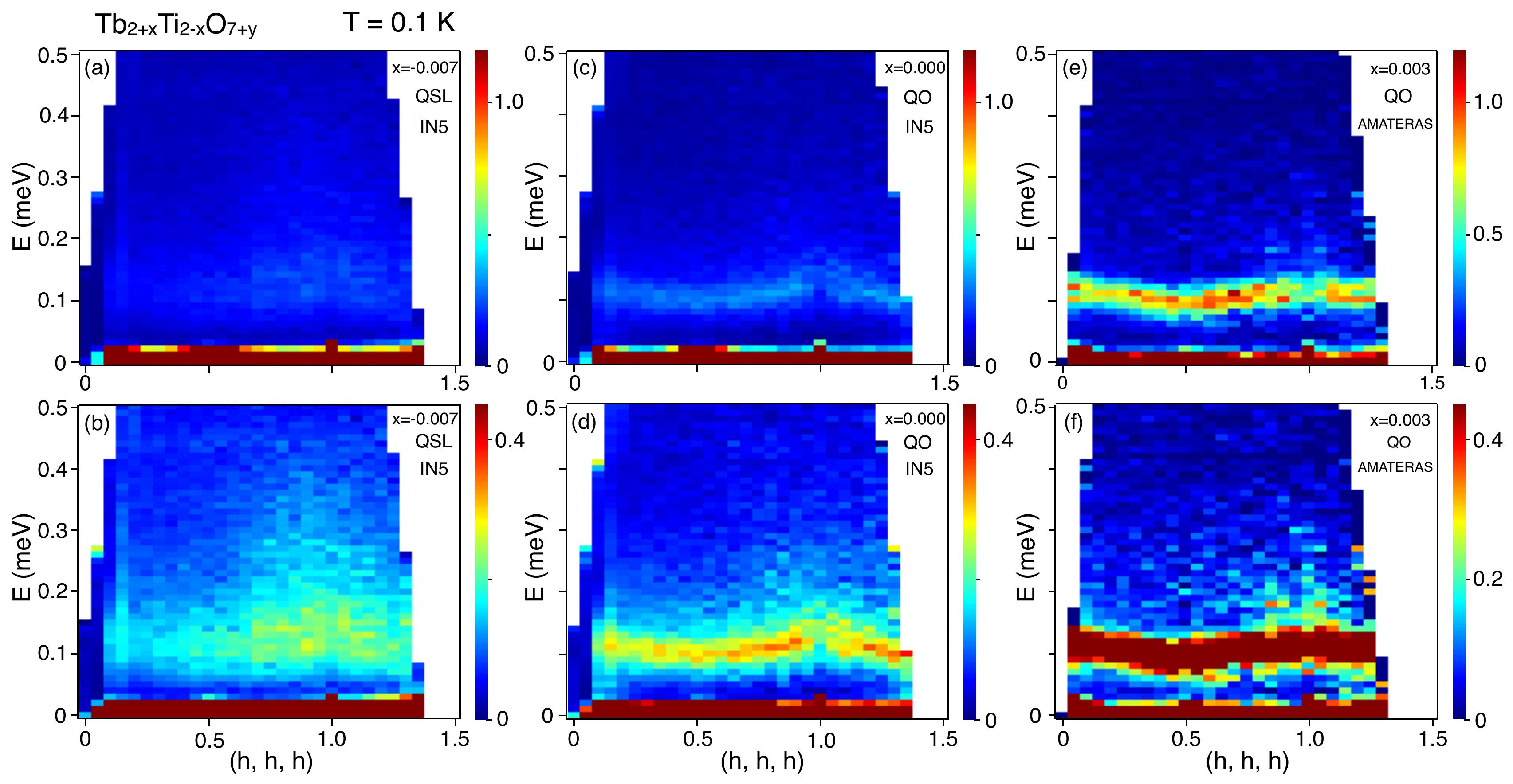}
\caption{(Color online) 
$Q$-$E$ slices along a [1,1,1] direction measured at $T=0.1$ K. 
The color bars show linear intensity scales of $S(\bm{Q},E)$ 
in the ``arb. units'' used in Fig.~\ref{Escan_Qav_0p1K_PhaseDiagram}. 
(a,b) show $S(\bm{Q},E)$ of the QSL sample with $x = -0.007$. 
(c,d) and (e,f) show $S(\bm{Q},E)$ of the QO samples with $x = 0.000$ and 0.003, respectively. 
Two colormaps of (a,c,e) and (b,d,f) are used for high and low intensity ranges, respectively. 
}
\label{EHHH_cut_0p1K}
\end{figure*}

Neutron scattering experiments for the $x = -0.007$ and 0.000 crystal samples 
were carried out on the time-of-flight (TOF) spectrometer IN5 \cite{Ollivier2011} 
operated with $\lambda = 8$ {\AA} at ILL. 
The energy resolution of this condition was $\Delta E = 0.021$ meV (FWHM) 
at the elastic position. 
The previous INS experiments \cite{Taniguchi13,Takatsu2016prl} using 
the powder samples ($x = -0.005$ and 0.005) were carried out on IN5 operated 
with $\lambda = 10$ {\AA}. 
The energy resolution of this condition was $\Delta E = 0.012$ meV (FWHM) 
at the elastic position. 
INS experiments for the $x = 0.003$ crystal sample were performed on the TOF 
spectrometer AMATERAS \cite{Nakajima2011} operated 
with $\lambda = 7$ {\AA} at J-PARC. 
The energy resolution of this condition was $\Delta E = 0.024$ meV (FWHM) at the elastic position. 
Each crystal sample was mounted in a dilution refrigerator 
so as to coincide its $(h,h,l)$ plane with 
the horizontal scattering plane of the spectrometer. 
The observed intensity data of the crystal samples were converted to $S(\bm{Q},E)$ 
using Lamp \cite{Richard1996} or Utsusemi \cite{Inamura2013}, 
and further corrected for absorption using a home-made program \cite{Kadowaki2018github}. 
Construction of four dimensional $S(\bm{Q},E)$ data object 
from a set of the TOF data taken by rotating each crystal sample
are performed using {H}{\footnotesize ORACE} \cite{Horace2016}. 

Background scattering of $S(\bm{Q},E)$ from sample holders etc. in the elastic channel 
was subtracted. 
Background in the inelastic channel, which is very small at least in $E<0$ at $T=0.1$ K, 
was neglected. 
To compare $S(\bm{Q},E)$ of the crystal samples 
we normalize $S(\bm{Q},E)$ using a relation 
$\int S(\bm{Q},E) d\bm{Q} dE = \text{const}$, where 
$\bm{Q}$ integrations extend to $|\bm{Q}|< \infty$ and $E$ is integrated 
in a range which covers all scattering contribution 
from the ground state doublet of the crystal field. 
This exact relation at $T = 0$ is 
approximated by that with $T = 0.1$ K and over a wide integration range: 
an $E$ range of $-0.1<E<0.5$ meV, and a $\bm{Q}$ range of 
$\bm{Q}=(h+k,h-k,l)$ with $0<h<0.9$, $0.75<l<1.75$, and $-0.25<k<0.25$ (r.l.u). 
This approximation is probably good for the present three crystal samples. 
To compare these data with $\int S(\bm{Q},E) d\bm{Q}$ of the powder samples \cite{Taniguchi13,Takatsu2016prl}, 
we used the same relation, where the integration range was replaced 
to $0.3 < |\bm{Q}|< 0.9$ {\AA}$^{-1}$ and $-0.1<E<0.5$ meV.

\section{Results and Discussion}
\subsubsection{Inelastic Neutron Scattering: Q-Averaged Spectra}
In Fig.~\ref{Escan_Qav_0p1K_PhaseDiagram} we show 
$\bm{Q}$-averaged $E$-scans ($E$-cuts) of $S(\bm{Q},E)$ at $0.1$ K 
for the crystal samples. 
In this figure, our previous results using the powder samples 
with $x=-0.005$ and 0.005 taken at $0.1$ K \cite{Taniguchi13,Takatsu2016prl} 
are also plotted for comparison. 
These data of the powder samples are $\bm{Q}$-averaged $S(\bm{Q},E)$ 
which are averaged in $0.3 < |\bm{Q}|< 0.9$ {\AA}$^{-1}$. 
To clearly show $x$ dependence of these $E$-cuts 
we approximately normalized the scattering intensities. 
By the present definition, 
the energy integrations of $[S(\bm{Q},E)]_{\bm{Q}\text{-average}}$ 
shown in Fig.~\ref{Escan_Qav_0p1K_PhaseDiagram} are the same for the five samples. 
These normalized ``arb. units'' are used in the subsequent figures. 

One can see from Fig.~\ref{Escan_Qav_0p1K_PhaseDiagram} 
that energy spectra of the two QSL samples 
with $x=-0.007$ and $-0.005$ show overall similarities. 
For the $x=-0.005$ sample, 
which is closer to the quantum phase transition, 
an increase of spectral weight at low energies ($0<E<0.06$ meV) 
are seen. 
As $x$ is increased to $x = 0.000$, which is in the QO range, 
the spectra are qualitatively changed; 
a small peak around $E=0.1$ meV appears in the $E$-cut. 
This excitation peak develops as $x$ is further increased to $x=0.003$ and $0.005$. 
We note that the energy spectra of the crystal and the powder samples 
show reasonable $x$-dependence, 
despite the fact that the crystal and powder samples were prepared in very different ways.  

The energy spectra obviously consist of two parts: 
energy-resolution-limited (nominally) elastic scattering 
and inelastic scattering. 
We note that the existence of these two scatterings have been commonly observed 
in INS experiments of all TTO samples, e.g. \cite{Yasui2002}, 
although $\bm{Q}$- and $E$-dependence of $S(\bm{Q},E)$ 
problematically depended on the sample quality. 
The energy-resolution limited scattering 
in the present experimental condition 
implies that two-spin time correlations 
$\langle S_i^z(0) S_j^z(t) \rangle$ have slowly decaying parts in 
a time scale much longer than 66 ps ($E \sim 0.01$ meV, 2.4 GHz, or 0.1 K). 
On the other hand, the inelastic scattering with the energy scale 0.1 meV 
implies that $\langle S_i^z(0) S_j^z(t) \rangle$ has a short time scale of 6.6 ps.

\subsubsection{Inelastic Neutron Scattering: Q-Dependent Spectra}
\begin{figure}
\centering
\includegraphics[width=8.0cm,clip]{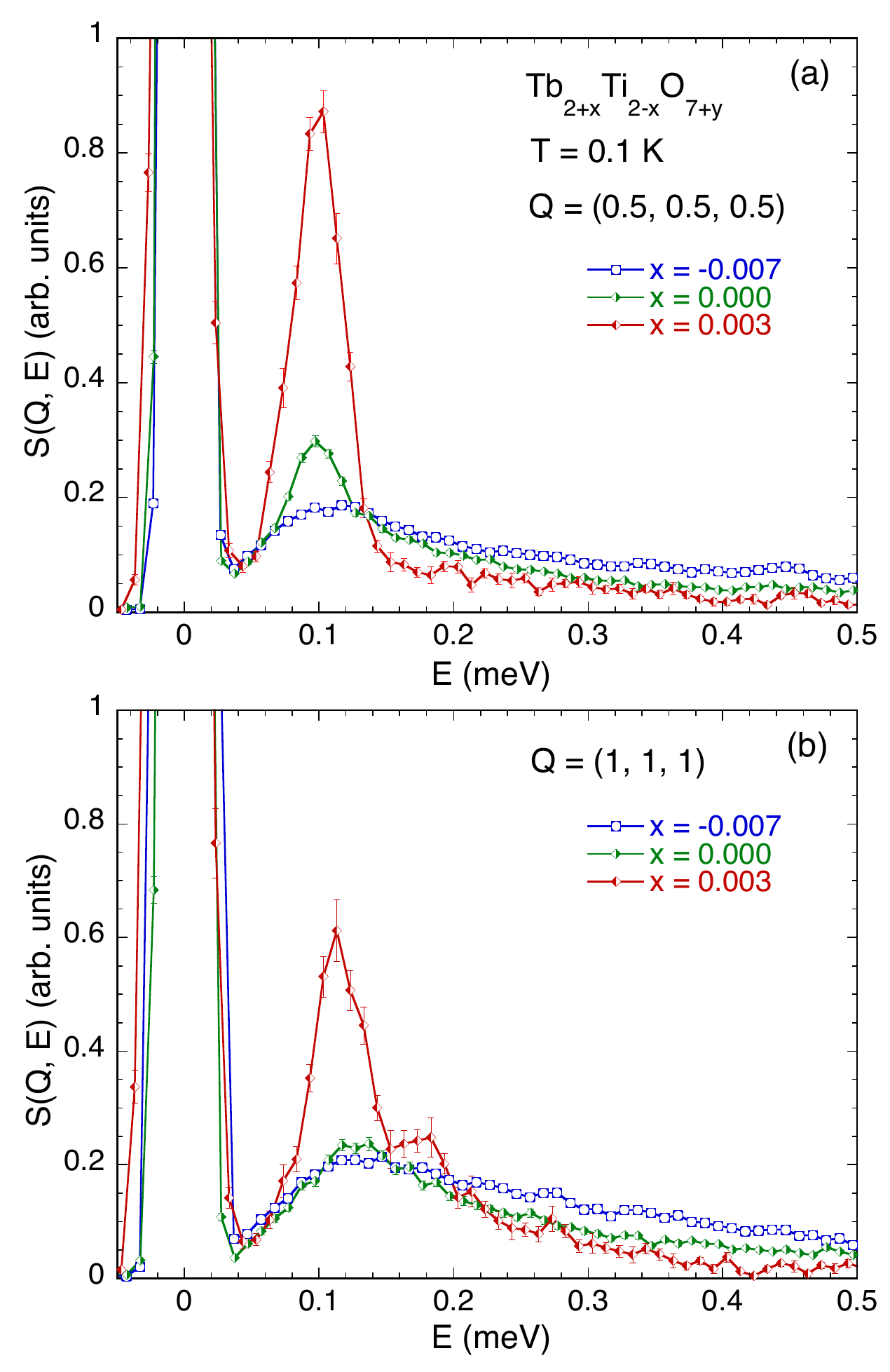}
\caption{(Color online) 
Inelastic energy spectra $S(\bm{Q},E)$ at $T=0.1$ K of the three samples 
are shown for (a) $\bm{Q}=(1/2,1/2,1/2)$ and (b) $\bm{Q}=(1,1,1)$. 
The ``arb. units'' of Fig.~\ref{Escan_Qav_0p1K_PhaseDiagram} are used. 
}
\label{Escan_0p1K_1_0p5key}
\end{figure}

Fig.~\ref{EHHH_cut_0p1K} shows typical $\bm{Q}$-dependence 
of the inelastic scattering at $T=0.1$ K 
by plotting $Q$-$E$ slices along a [1,1,1] direction. 
For closer inspection of energy spectra, 
$E$-cuts at $\bm{Q} = (1/2,1/2,1/2)$ and $(1,1,1)$ 
are plotted in Fig.~\ref{Escan_0p1K_1_0p5key}. 
One can see from these figures that the QSL sample shows 
gapless continuum excitation spectra [Figs.~\ref{EHHH_cut_0p1K}(a) and (b)]. 
These continuum excitations may possibly be 
those of ``magnetic monopole'' \cite{Hermele04,Chen2017} 
(or vison \cite{Gingras14}) of the U(1) QSL state. 
On the other hand, 
for the QO sample with $x = 0.000$ [Figs.~\ref{EHHH_cut_0p1K}(c) and (d)]
gapped pseudospin wave excitation 
with flat dispersion relation emerges from the continuum excitation. 
This pseudospin wave excitation becomes more distinct 
for the QO sample with $x = 0.003$ [Figs.~\ref{EHHH_cut_0p1K}(e) and (f)]. 
We also searched $S(\bm{Q},E)$ for the propagating excitations 
reported in Ref.~\cite{Guitteny2013}, although 
nothing similar was found in the present data. 

The pseudospin wave excitation can be understood as 
composite wave of magnetic-dipole and electric-quadrupole moments 
as discussed in Refs.~\cite{Kadowaki2015,Takatsu2016prl}.  
We note that this excitation bears resemblance to 
spin-orbital excitations in $d$-electron 
systems \cite{Khaliullin2005,Tomiyasu2006,Helme2009,Das2018}.
Starting from the proposed model parameter \cite{Takatsu2016prl},
we are now performing refinements of interaction parameters 
to reproduce the observed dispersion relation of the $x = 0.003$ sample. 
In addition to this pseudospin wave, 
there remains the continuum excitation in Figs.~\ref{EHHH_cut_0p1K}(e) and (f). 
Especially, the E-cut at $\bm{Q} = (1,1,1)$ [Fig.~\ref{Escan_0p1K_1_0p5key}(b)] 
shows that the continuum excitation ($E>0.15$ meV) is substantially strong. 
These facts suggest that the QO state has large zero-point quantum fluctuations. 

Finally, let us make a comment on a simple question of 
``what the off-stoichiometry $x$ does?'' 
in relation to the phase diagram of Fig.~\ref{Escan_Qav_0p1K_PhaseDiagram}(a). 
A possible answer to this question or a scenario is as follows. 
In the $x=0$ sample, there is a tiny amount of site exchange 
between non-magnetic Ti$^{4+}$ and magnetic Tb$^{3+}$ sites. 
As a consequence, there remains randomness effects in the nominally stoichiometric sample. 
As $x$ is increased from $x=0$, more magnetic Tb$^{3+}$ or Tb$^{4+}$ ions 
occupy the Ti$^{4+}$ site. 
These magnetic ions reduce the effective magnetic interactions, 
while quadrupole interactions are very weakly affected. 
For small $x>0$, the QO state is observed. 
By further increasing $x$ the randomness effects 
on quadrupole interactions become stronger and 
the quadrupole LRO disappears ($x>0.04$). 
On the other hand, as $x$ is decreased from $x=0$, 
magnetic Tb$^{3+}$ ions residing on the Ti$^{4+}$ site are removed. 
For $x < -0.0025$ the effective magnetic interactions become 
sufficiently strong and the QSL state is observed. 
Recently randomness effects on QSL states are studied theoretically 
\cite{Kawamura2014,Kawamura2017,Savary2017PRL,Benton2017,Zhu2017} 
and experimentally \cite{Wen2017,Martin2017,Sibille2017NC}. 
In these studies, much larger model disorders than those in TTO ($-0.01<x<0.005$) 
are studied.

\section{Conclusions}
The ground states of the frustrated pyrochlore oxide Tb$_{2+x}$Ti$_{2-x}$O$_{7+y}$ 
have been studied by inelastic neutron scattering experiments using 
three single-crystal samples: 
one putative QSL sample with $x=-0.007$ ($<x_{\text{c}}$) 
and 
two QO samples with $x=0.000$ and 0.003. 
Small concentration gradient ($|\Delta x | < 0.002$) of these samples 
has enabled us to observe $x$-dependence 
of inelastic excitation spectra at low temperatures. 
The QSL sample shows continuum excitation spectra with an energy scale 0.1 meV 
as well as energy-resolution-limited (nominally) elastic scattering. 
As $x$ is increased, 
pseudospin wave of the QO state emerges from the continuum excitation. 
\begin{acknowledgments}
This work was supported by JSPS KAKENHI grant numbers 25400345 and 26400336. 
The neutron scattering performed using ILL IN5 (France) 
was transferred from JRR-3M HER (proposal 11567, 15545) 
with the approval of ISSP, Univ. of Tokyo, and JAEA, Tokai, Japan. 
The neutron scattering experiments at J-PARC AMATERAS were carried out 
under a research project number 2016A0327.
\end{acknowledgments}

\section{Appendix*}
The off-stoichiometry parameters ($x$) of Tb$_{2+x}$Ti$_{2-x}$O$_{7+y}$ samples 
were evaluated by measuring the lattice parameter 
of small crystals ($\sim 1$ mm) cut from single-crystal rods \cite{Wakita2016}. 
We performed high-resolution $\theta$-$2\theta$ scans on 
powder mixtures ($\sim 5$ mg) of polycrystalline Si and 
crushed-crystalline Tb$_{2+x}$Ti$_{2-x}$O$_{7+y}$ using 
the X-ray diffractometer. 
In these measurements, temperature of the mixture was controlled 
at $T=26.0 \pm 0.1^{\circ} \text{C}$ to minimize experimental error. 
We determined $x$ using the relation $a(T=26.0^{\circ} \text{C},x)=0.124418x+10.15280$ \cite{Wakita2016}, 
which was measured using powder samples \cite{Taniguchi13}.

\begin{figure}
\centering
\includegraphics[width=8.0cm,clip]{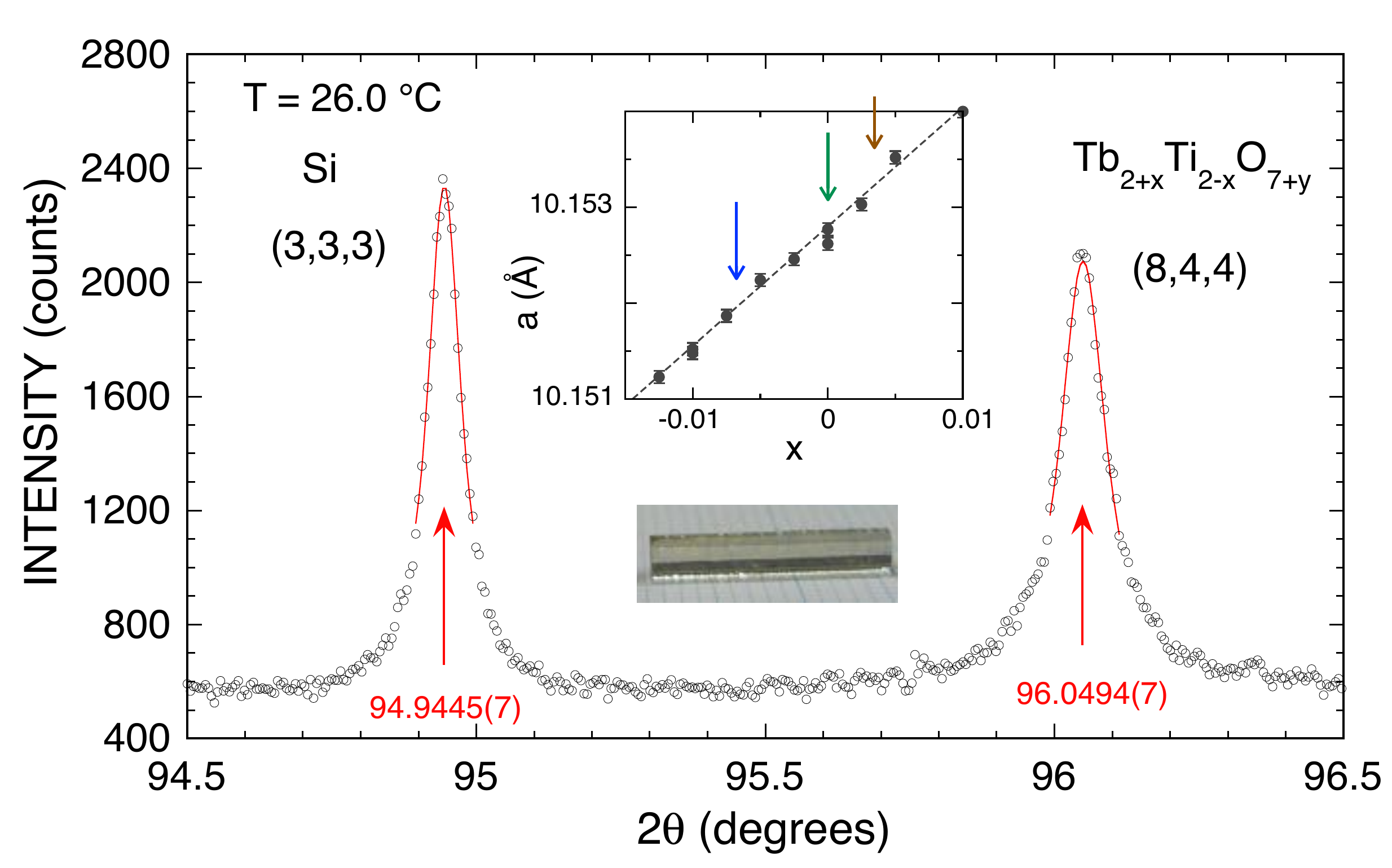}
\caption{(Color online) 
A typical $\theta$-$2\theta$ scan of a powder mixture 
of Si and crushed-crystalline Tb$_{2+x}$Ti$_{2-x}$O$_{7+y}$. 
The inset plot shows the lattice parameter $a(T=26.0^{\circ} \text{C},x)$ of 
Ref.~\cite{Wakita2016}, where three arrows indicate the three samples 
for the present single-crystal INS experiments. 
The inset photograph shows a 15 mm crystal rod, one of 
the QSL multi-crystal sample. 
}
\label{theta2theta_scan}
\end{figure}

Fig.~\ref{theta2theta_scan} shows 
a typical $\theta$-$2\theta$ scan carried out to measure 
$a$ and to evaluate $x$ of a small crystal sample. 
It should be noted that although there are only two peaks of Si 333 and 
TTO 844 reflections in the scan range [$94.5< 2\theta < 96.5$ (deg)], 
HEIDENHAIN absolute angle encoders installed in the SmartLab 
ensure high reproducibility of the result. 
This reproducibility is seen 
in Fig.~3 of Ref.~\cite{Goto12} and Fig.~6 of Ref.~\cite{Takatsu2014}. 
In order to further reduce slight systematic errors of $2\theta$, 
which are brought about possibly by occasional optical alignments, 
we repeated the same $\theta$-$2\theta$ scan on a standard sample 
and made a correction. 
This standard sample was a mixture of Si and the $x=-0.0075$ TTO powder \cite{Taniguchi13}. 
Thus, we note that 
our evaluation of the lattice parameter depends on both this standard sample and 
the certified lattice parameter $a=$5.43123(8) {\AA} of Si (NIST SRM640d) \cite{SRM640d}. 
In addition, we note that the oxidizing condition determining $y$ also affects the $x$ evaluation. 

We would like to finally mention another problem of the TTO preparation 
caused by the starting material Tb$_{4}$O$_{7}$. 
Since Tb$_{4}$O$_{7}$ is not a pure chemical substance, 
it should be expressed probably as Tb$_{4}$O$_{7+\delta}$ with small $\delta$. 
This $\delta$, normally unknown, depends on details of the production process of 
each chemical company. 
The starting material Tb$_{4}$O$_{7}$ used in our investigations from 2012, 
i.e., Refs.~\cite{Taniguchi13,Kadowaki2015,Takatsu2016prl,Takatsu2016JPCS,Wakita2016,Takatsu2017} 
and the preset study, 
was from a single batch produced by Shin-Etsu Chemical. 

Because of these technical problems, 
one has to be very cautious to compare 
evaluations of the composition $x$ or/and $y$ of TTO samples 
by different investigation groups and even by the same group. 
We think that our samples are self-consistent among 
Refs.~\cite{Taniguchi13,Kadowaki2015,Takatsu2016prl,Takatsu2016JPCS,Wakita2016,Takatsu2017} 
and the present study. 
The sample used in Ref.~\cite{Takatsu12}, however, 
shows slight breaking of this self-consistency, 
which is seen in Fig.~2 of Ref.~\cite{Taniguchi13}. 
Thus we think that it is not sensible to make comments here 
on samples of other investigation groups. 

\bibliography{TTO_HK_pdflatex}
\end{document}